\documentclass[apjl]{emulateapj}
\usepackage[dutch, english]{babel}
\usepackage{amssymb}
\usepackage{amsmath}
\usepackage{mathrsfs}
\usepackage{latexsym}
\usepackage{longtable}
\usepackage{epsf}
\usepackage{color}
\usepackage{graphicx}
\usepackage{float}
\usepackage{enumerate}

\shorttitle{08576nr292: Discovery of a disk-jet system}
\shortauthors{L.E. Ellerbroek et al.}

\begin{document}

\title{The Intermediate-mass Young Stellar Object 08576nr292: \\Discovery of a
  disk-jet system\altaffilmark{$\bigstar$}}

\author{Lucas~E.~Ellerbroek\altaffilmark{1}}
\author{Lex~Kaper\altaffilmark{1}}
\author{Arjan~Bik\altaffilmark{2}}
\author{Alex~de~Koter\altaffilmark{1,4}}
\author{Matthew~Horrobin\altaffilmark{5}}
\author{Elena~Puga\altaffilmark{3,6}}
\author{Hugues~Sana\altaffilmark{1}}
\author{Laurens~B.F.M.~Waters\altaffilmark{7,1}}
\email{l.e.ellerbroek@uva.nl}

\altaffiltext{1}{Sterrenkundig Instituut Anton Pannekoek, University
  of Amsterdam, Science Park 904, P.O. Box 94249, 1090 GE Amsterdam,
  The Netherlands}
\altaffiltext{2}{Max-Planck-Institut f\"{u}r Astronomie,
  K\"{o}nigstuhl 17, 69117 Heidelberg, Germany} 
\altaffiltext{3}{Instituut voor Sterrenkunde, Celestijnenlaan 200D, B-3001 Leuven, Belgium}
\altaffiltext{4}{Astronomical Institute, Utrecht University,
  Princetonplein 5, 3584CC Utrecht, The Netherlands}
\altaffiltext{5}{I.\ Physikalisches Institut, Universit\"{a}t zu K\"{o}ln, 50937 K\"{o}ln, Germany}
\altaffiltext{6}{Centro de Astrobiolog\'{i}a (CSIC-INTA), 28850
  Torrej\'{o}n de Ardoz, Madrid, Spain}
\altaffiltext{7}{SRON, Sorbonnelaan 2, 3584 CA, Utrecht, The Netherlands}
\altaffiltext{$\bigstar$}{Based on observations performed with X-shooter (program P84.C-0604) and SINFONI (program P78.C-0780)
  mounted on the ESO {\it Very Large Telescope} on Cerro Paranal,
  Chile}

\begin{abstract}
We present observations of the embedded massive young stellar object (YSO) candidate 08576nr292, obtained with X-shooter and SINFONI on the ESO {\it Very Large Telescope} (VLT). The flux-calibrated, medium-resolution X-shooter spectrum (300~--~2500~nm) includes over 300 emission lines, but no (photospheric) absorption lines and is consistent with a reddened disk spectrum. Among the emission lines are three hydrogen series and helium lines, both permitted and forbidden metal lines, and CO first-overtone emission. A representative sample of lines with different morphologies is presented. The H$\alpha$ and Ca~{\sc ii} triplet lines are very strong, with profiles indicative of outflow and -- possibly -- infall, usually observed in accreting stars. These lines include a blue-shifted absorption component at $\sim -125$~km~s$^{-1}$. The He~{\sc i} and metal-line profiles are double-peaked, with a likely origin in a circumstellar disk. The forbidden lines, associated with outflow, have a single blue-shifted emission component centered at   $-125$~km~s$^{-1}$, coinciding with the absorption components in H$\alpha$ and Ca~{\sc ii}. SINFONI H- and K-band integral-field spectroscopy of the cluster environment demonstrates that the [Fe~{\sc ii}] emission is produced by a jet originating at the location of 08576nr292. Because the spectral type of the central object cannot be determined, its mass remains uncertain. We argue that 08576nr292 is an intermediate-mass YSO with a high accretion rate ($\dot{M}_{\rm acc} \sim 10^{-5}$ -- 10$^{-6}$~$M_\odot$~yr$^{-1}$). These observations demonstrate the potential of X-shooter and SINFONI to study in great detail an accretion disk-jet system, rarely seen around the more massive YSOs.
\end{abstract}

\keywords{stars: formation --- stars: mass-loss --- stars:
  pre-main sequence --- ISM: jets and outflows}
  
\received{February 21, 2011}
\accepted{March 22, 2011}  
  
\section{Introduction}

The formation process of massive stars is poorly understood (for a recent review, see \citealt{Zinnecker:2007p1102}). Besides complicated theoretical aspects, observations of forming massive stars are limited due to their short formation timescale, the strong obscuration of their birth places, and their relatively small number. A current key question is whether massive stars form through accretion in a way similar to low- and intermediate-mass stars (e.g. \citealt{Shu:1977p1819, Palla:1993p1826, Yorke:2002p1824, Kuiper:2010p1883}), or that radiation pressure and stellar wind prevent the formation of massive stars, calling for alternative formation scenarios such as stellar mergers in dense and young clusters (\citealt{Bonnell:1998p1912, Baumgardt:2010p1889}). The detection of the signatures of formation, i.e. circumstellar disks and collimated jets, around massive young stellar objects (MYSOs) would provide an important step in addressing this question.
 
Up to now observational evidence for the presence of disks and/or collimated outflows has been secured for about a dozen of (candidate) MYSOs (e.g. \citealt{Cesaroni:2007p1925, Sandell:2010p1946, Kraus:2010p987, Quanz:2010p1974, Guzman:2010p2010, Zapata:2010}). The current mass of these objects is often unclear, and their identification as a MYSO is mainly based on their luminosity. Most of them have a corresponding ZAMS mass less than 20~M$_{\odot}$ (i.e. spectral type later than O8). 

We have initiated an observing campaign to extend the spectral coverage of a sample of MYSO candidates as far to the blue as possible, to better determine their photospheric properties, to study the onset of stellar winds, and to characterize the physical structure of their accretion disks. A sample of MYSO candidates (\citealt{Bik:2006p4}, hereafter B06) and young OB stars (\citealt{Bik:2005p14}, hereafter B05) has been obtained from a near-infrared survey of southern star-forming regions including ultra-compact H{\sc ii} (UCH{\sc ii}) regions (\citealt{Bik:2004p2346}). Among these is the region RCW~36, centered on the IRAS point source 08576-4334, with the characteristic infrared colors of an UCH{\sc ii}.

This point source is associated with Vela Cloud~C (\citealt{Yamaguchi:p1168}), for which the estimated distance is $0.7 \pm 0.2$~kpc (\citealt{Liseau:1992p2550}). The kinematic distance, however, from the rotation curve of \cite{Brand:1993p2751} and a Local Standard of Rest velocity ($v_{LSR}$) of 7.5~km~s$^{-1}$ (\citealt{Bronfman:1996}) is 2.2~kpc. In this work a distance of 0.7~kpc is adopted, consistent with the spectroscopic parallax obtained for OB stars in the cluster (B05). This region includes the MYSO candidate 08576nr292. Based on J-K photometry, B06 estimate the spectral type of the central object at mid-B. 08576nr292 has a strong infrared excess, and its K-band spectrum includes a broad Br$\gamma$ emission line and CO band-head emission, characteristic features of MYSOs (cf. \citealt{Hanson:1997p55, Blum:2004p89}). The CO band-head emission can be modeled by a Keplerian rotating, circumstellar disk (\citealt{Bik:2004p180}). B06 propose that such a disk likely is a remnant accretion disk. \cite{Wheelwright:2010p56} derive an inclination angle of $i=17.8^{\circ+0.8}_{-0.4}$ based on the CO-profiles, which means that the disk is seen almost pole-on.

In the following we present the optical to near-infrared spectrum of 08576nr292 obtained with the new spectrograph X-shooter, as well as near-infrared integral-field spectroscopy of its environment with SINFONI, both mounted on the ESO {\it Very Large Telescope}  (VLT) at Paranal, Chile. The combination of these data sets reveals important new information on the circumstellar environment of this MYSO candidate. 

\section{Observations and data reduction}

\subsection{VLT/X-shooter}

\begin{figure*}[t]
  \center{\includegraphics[width=18cm]{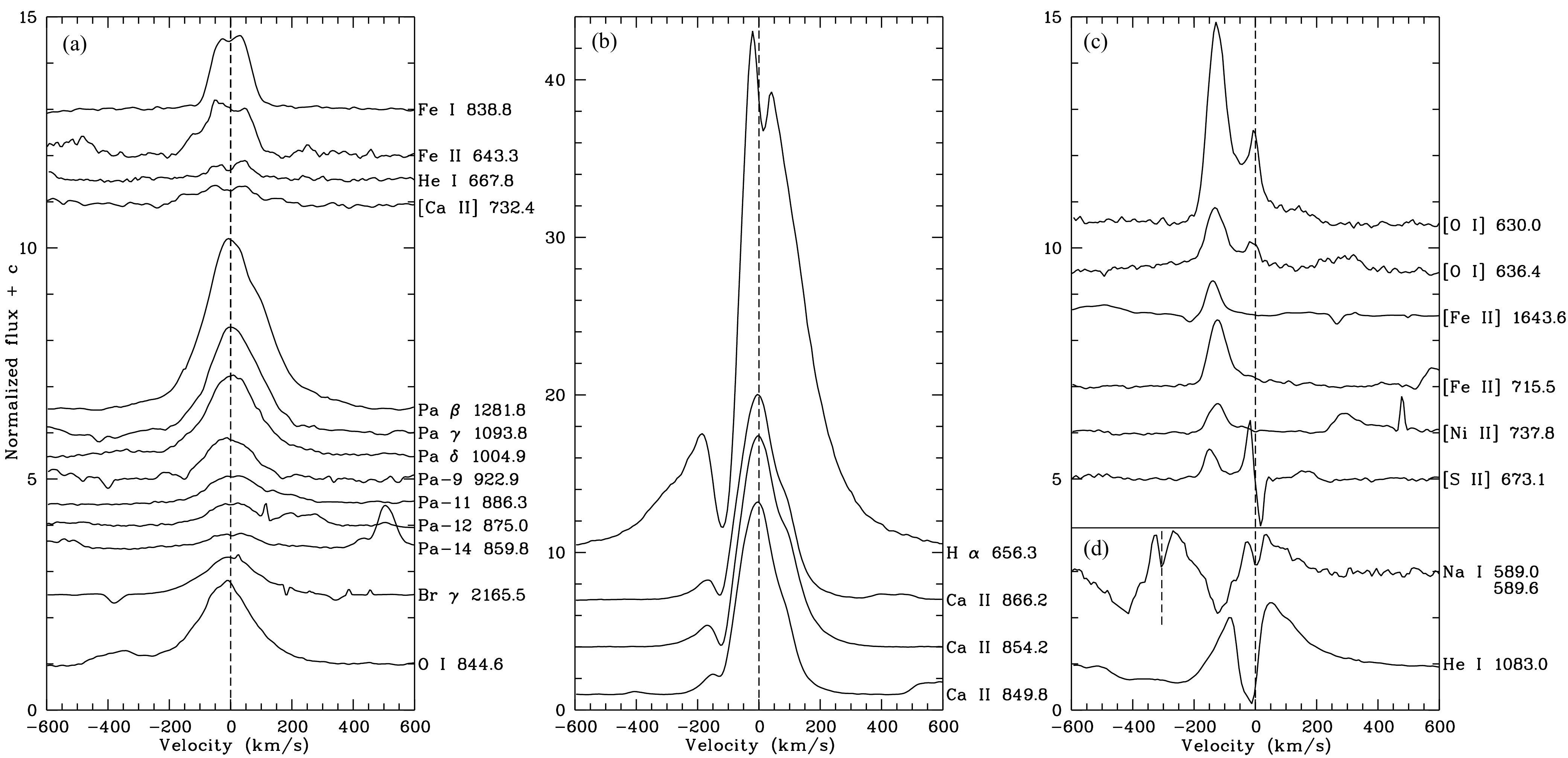}}
  \caption{A representative sample of the 300+ normalized spectral lines and their various profiles, detected by X-shooter in observation III of 08576nr292. (a)~Double-peaked and single-peaked lines; (b)~Emission lines with a blue-shifted absorption component at $-125$~km~s$^{-1}$ and a `red shoulder' (Ca~{\sc ii}). The central absorption in H$\alpha$ is due to nebular contamination; (c)~Forbidden transitions blue-shifted at $-125$~km~s$^{-1}$; (d)~Peculiar profiles. The spectra are normalized. The rest wavelengths (nm) are in air.}
\label{fig:lineprofiles}
\end{figure*}

The MYSO candidate 08576nr292 (RA (2000.0) $08^{\rm   h}59'21.7''$; Dec (2000.0) $-43^\circ45'31.0''$) was observed as part of the X-shooter GTO program ``Probing the earliest evolutionary phases of the most massive stars'' in February 2010. A selection of the most important spectral features is presented in Fig.~\ref{fig:lineprofiles}. Each observation comprised two separate exposures at different nodding positions on the slit (see Table~\ref{tab:obs}). In the NIR arm the exposure time was split into detector integration times (DITs) of 50s each.

\begin{table}[!t]
\begin{center}
\begin{tabular}{|c|c|c|c|c|}
\hline
Obs.\# & date & exp.time & nod throw & pos. angle\\
\hline
I & 22-02-10 UT $5^{\rm h}06$ & 2$\times$900 s & 2'' & 19.52$^\circ$ (1)\\
II & 22-02-10 UT $5^{\rm h}36$ & 2$\times$900 s & 2'' &  19.52$^\circ$ (1)\\
III & 23-02-10 UT $4^{\rm h}50$ & 2$\times$600 s & 5'' & 50.26$^\circ$ (2)\\
\hline
\end{tabular}
\caption{{\rm List of the observations carried out with X-shooter. The position angle is the slit angle on the sky from north to east. The numbers between parentheses refer to the slit orientations depicted in Fig.~\ref{fig:sinfoni}.}}
\label{tab:obs}
\end{center}
\end{table}

The X-shooter wavelength range is 300--2500~nm, split in three arms using dichroics: UVB ($300-560$ nm), VIS ($550-1020$ nm) and NIR ($1000-2500$ nm). For a detailed description of the instrument, see \cite{DOdorico:2006p2081}. Given the strong reddening of the source, different slit widths were used in the UVB (1.0$''$), VIS (0.9$''$) and NIR (0.4$''$) arms. The resolving power was calculated from wavelength calibration frames and averages to 5100, 8800, and 11300 (corresponding to 59, 35 and 27 km~s$^{-1}$ per resolution element) in the three arms, respectively. The V-band seeing varied between 0.6$''$ and 0.9$''$ during the observations. A telluric standard star (HD80055, A0V) was observed immediately after 08576nr292. The spectrophotometric standard star GD71, a DA white dwarf, was observed in twilight.

The ESO X-shooter pipeline version 1.0.0 (\citealt{Goldoni:2006p3452, Modigliani:2010p3073}) was used to obtain reduced 2D spectra.  From these 2D-spectra, 1D-spectra were extracted and normalized to the continuum using our own routines. For the flux calibration we used a flux table for GD71 in the range 300 -- 1000 nm, and scaled a Rayleigh-Jeans function to the 2MASS magnitudes for HD80055 in the range 1000 -- 2500 nm. We estimated slit losses based on the seeing conditions. The derived J-, H- and K-band fluxes are within 30\% of the 2MASS values: $\textrm{J} = 11.82 \pm 0.03, \textrm{H} = 10.38 \pm 0.04, \textrm{K} = 9.32 \pm 0.04$.

\subsection{VLT/SINFONI}
\label{sec:sinfoniobs}

Observations of the cluster environment of 08576nr292 were made with the integral field spectrograph SINFONI (\citealt{Eisenhauer:2003p2117, Bonnet:2004p2130}). It produces an H- and K-band spectrum at every pixel ($0.125 \arcsec ” \times 0.125 \arcsec$) in its field-of-view (8\arcsec $\times$ 8\arcsec) at R $\sim$ 1,500. The SINFONI data were collected on March 5, 7 and 19, 2007. Each observation was accompanied by the observation of an early B-type standard star observed immediately after the science observations, and used for flux calibration and telluric absorption correction. 

Different pointings in a raster pattern were used to cover a large area surrounding 08576nr292 (for more details, see \citealt{Bik:2010p3}). The DIT was 30~s per pointing. With every point being covered twice, this results in a total on-source integration time of 60~s.  Sky frames were taken every 3 minutes using an empty area on the sky. The K-band seeing was 0.4$''$ for the first two observing nights (part A in Fig.~\ref{fig:sinfoni}). During the last night (part B), the seeing was  0.7$''$. The south west corner of the image was not observed.

The data were reduced using the SINFONI {\sc SPRED} pipeline (version 1.37) developed by the MPE SINFONI consortium (\citealt{Schreiber:2004p2146, Abuter:2006p2164}). Velocity maps were created by fitting a gaussian profile to every spatial pixel in the data cubes. The velocities were converted to the  LSR.

\section{Results}

\subsection{X-shooter spectrum of 08576nr292}
\label{sec:xshooterres}

In the X-shooter spectra, the signal to noise ratio (SNR) is about 80 at 1200 nm and 50 at 650 nm, respectively, and less than 10 below 550 nm due to interstellar extinction. Despite the high SNR, no photospheric absorption lines are detected. The spectrum contains over 300 emission lines, almost all of which could be identified. A representative sample of these lines highlighting the different morphologies is displayed in Fig.~\ref{fig:lineprofiles}. The velocity scale is corrected for the Earth's motion with respect to the LSR. 

The hydrogen Balmer, Paschen and Brackett series are in emission, as well as many different metal lines. Close to half of all the identified lines are iron lines. Some variation in the peak-to-peak flux ratio was observed in some Fe~{\sc i} lines between observations I/II and III. 

{\it (i) Double- and single-peaked emission lines:} The 230 permitted emission lines are predominantly from H~{\sc i}, He~{\sc i}, N~{\sc i}, O~{\sc i}, Mg~{\sc i}, Si~{\sc i}, Ni~{\sc i}, Fe~{\sc i} and Fe~{\sc ii}. About 50 of these lines (such as most He~{\sc i} and Fe~{\sc i} lines) are double peaked with a peak separation of 60 to 100~km~s$^{-1}$ (Fig.~\ref{fig:lineprofiles}a). The double-peaked signature is consistent with formation in a Keplerian rotating circumstellar disk, as suggested by the modelling of the CO band-heads at 2.3~$\mu$m present in the spectrum (\citealt{Bik:2004p180}).

{\it (ii) Emission lines with signatures of outflow and infall:} There are a number of lines with strong, single peaks centered around the systemic velocity. Examples of these are the lower transitions in the Paschen and Brackett series, and the O~{\sc i} lines (Fig.~\ref{fig:lineprofiles}a). By far the strongest lines are H$\alpha$ and the Ca~{\sc ii} triplet (Fig.~\ref{fig:lineprofiles}b), which include a narrow absorption feature, blue-shifted to $- 125\pm 6$~km~s$^{-1}$, a signature of outflow. The Ca~{\sc ii} lines have a red shoulder, a possible signature of infall onto the central star. A similar profile is seen in Pa$\beta$. Note that the apparent double peak in H$\alpha$ is an artefact of the subtraction of the nebular component, which varied over the slit.

{\it (iii) Forbidden lines:} A total of 67 identified forbidden lines are detected, 46 of which are [Fe~{\sc ii}]; a representative sample is displayed in Figure~\ref{fig:lineprofiles}c. These lines have a single peak centered at a blue-shifted velocity of $-125 \pm 8$~km~s$^{-1}$. This is the same velocity shift as that of the absorption features in H$\alpha$ and Ca~{\sc ii}. The peaks are slightly asymmetric with an extended wing towards the red and a sharp blue cutoff. The [O~{\sc i}] and [S~{\sc ii}] lines have a very weak red-shifted counterpart at $+140 \pm 15$~km~s$^{-1}$. The [O~{\sc i}] doublet lines at 630~nm are the strongest of the forbidden transitions. These lines, as well as the [S~{\sc ii}] lines, include a component at a low blueshifted velocity ($\sim - 10$~km~s$^{-1}$), that may be associated with a hot wind close to the star (cf. \citealt{Hartigan:1995p1041}). 

Forbidden line emission is produced by collisionally excited ions in a low-density medium. In this case, the two components at blue- and redshifted radial velocities suggest a bipolar outflow. The [Ca~{\sc ii}] doublet at 729~/~732 nm (Fig.~\ref{fig:lineprofiles}a) are the only detected forbidden lines that do not have a blue offset, but a double-peaked signature.

{\it (iv) Peculiar profiles:} There are two peculiar line profiles that do not fit in any of the above descriptions (Fig.~\ref{fig:lineprofiles}d). The first is the Na~{\sc i} $\lambda$589~nm doublet, which includes an interstellar absorption feature, and a blue-shifted absorption component centered at $-125$~km~s$^{-1}$. This absorption component is broader than that seen in Ca~{\sc ii} and H$\alpha$. Another peculiar profile is seen in the He~{\sc i} $\lambda$1083~nm line, which has a deep and narrow central absorption component and a broad blue-shifted absorption trough extending from $-200$~up~to~$-450$~km~s$^{-1}$.

\begin{figure*}[t]
\center{\includegraphics[width=18cm]{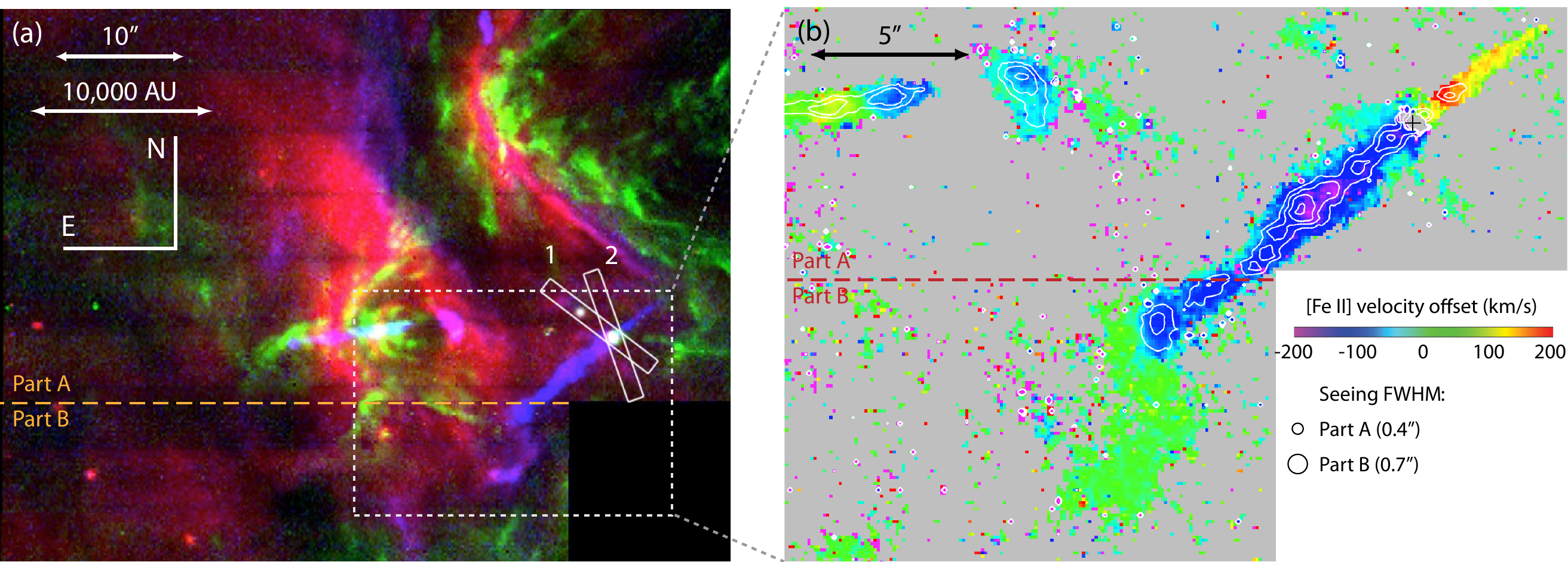}}
\caption{(a) Detail of the SINFONI image of IRAS 08576-4334 (size: $57''\times43''$). Color channels trace: H$_2$ (green), Br$\gamma$ (red) and [Fe~{\sc ii}]   (blue). The results from the different observing nights, as described in section~\ref{sec:sinfoniobs}, are separated by the orange dotted line. The X-shooter slit orientations are depicted (slit size: $1''\times 11''$). (b) Velocity offset map (size: $25''\times18''$) of the [Fe~{\sc ii}] $\lambda$1644~nm line. The color gradient corresponds to the offset of the peak with respect to the $v_{\rm LSR}$. Grey color represents areas where no [Fe~{\sc ii}]  emission was detected. The white contours represent the line flux, with logarithmic contour values between 1.0$\times 10^4$ and 2.3$\times 10^5$ erg s$^{-1}$ cm$^{-2}$ $\mu$m$^{-1}$ sr$^{-1}$. The black cross indicates the position of 08576nr292, where [Fe II] emission was blended with Br-12; this area is also colored grey. We note the presence of another jet system in the NE corner of the field of view.}
\label{fig:sinfoni}
\end{figure*}

\subsection{SINFONI data}

In Figure~\ref{fig:sinfoni}a, a line map is plotted of the environment of 08576nr292 with tracers of ionized gas (Br$\gamma$~$\lambda$2166~nm in red), shocked or UV excited gas (H$_2$~$\lambda$2121~nm in green) and low-density gas ([Fe~{\sc ii}]~$\lambda$1644~nm in blue). A collimated bipolar outflow, from now on referred to as `jet', is clearly detected in [Fe~{\sc ii}], originating at 08576nr292. The jet is also detected in Br$\gamma$ with a similar velocity structure.

The velocity map (Fig.~\ref{fig:sinfoni}b) shows that both lobes of the jet are moving away from the source. The velocity of the blue-shifted lobe corresponds to the peak velocity of the forbidden lines and the absorption features in H$\alpha$ and Ca~{\sc ii}. In both lobes, some regions peak up to 200~km~s$^{-1}$. Relatively high velocities are reached in those parts of the jet where the flux is relatively high (indicated by white contours). Careful inspection, however, reveals that the velocity peaks slightly closer to the source than does the flux. This suggests that fast-moving material is ploughing into slower moving material. This is commonly seen in bow-shocked jets that reveal an episodic accretion history (\citealt{Raga:1990p3264, Arce:2007p2643}).

It is clear that the acceleration of the jet material occurs close to the star ($< 0.5'' \equiv 350$~AU). The projected lengths of the lobes are 9,000~AU (blue-shifted south-east lobe) and 3,000~AU (red-shifted north-west lobe). Practically no H$_2$ emission is detected throughout the jet, but this might be due to the short exposure time. 

The jet does not seem to end in a bow-shock. The end of the receding (NW) lobe may be obscured by foreground dust and at the end of the approaching (SE) lobe the [Fe~{\sc ii}] emission changes velocity (the green-colored area in Fig.~\ref{fig:sinfoni}b). It is unclear whether this material is related to the jet, or part of another structure in the ISM. 

\section{Discussion}
\subsection{Evidence for a disk-jet system}
\label{sec:evidence}

\begin{figure}[t]
\center{\includegraphics[height=8cm]{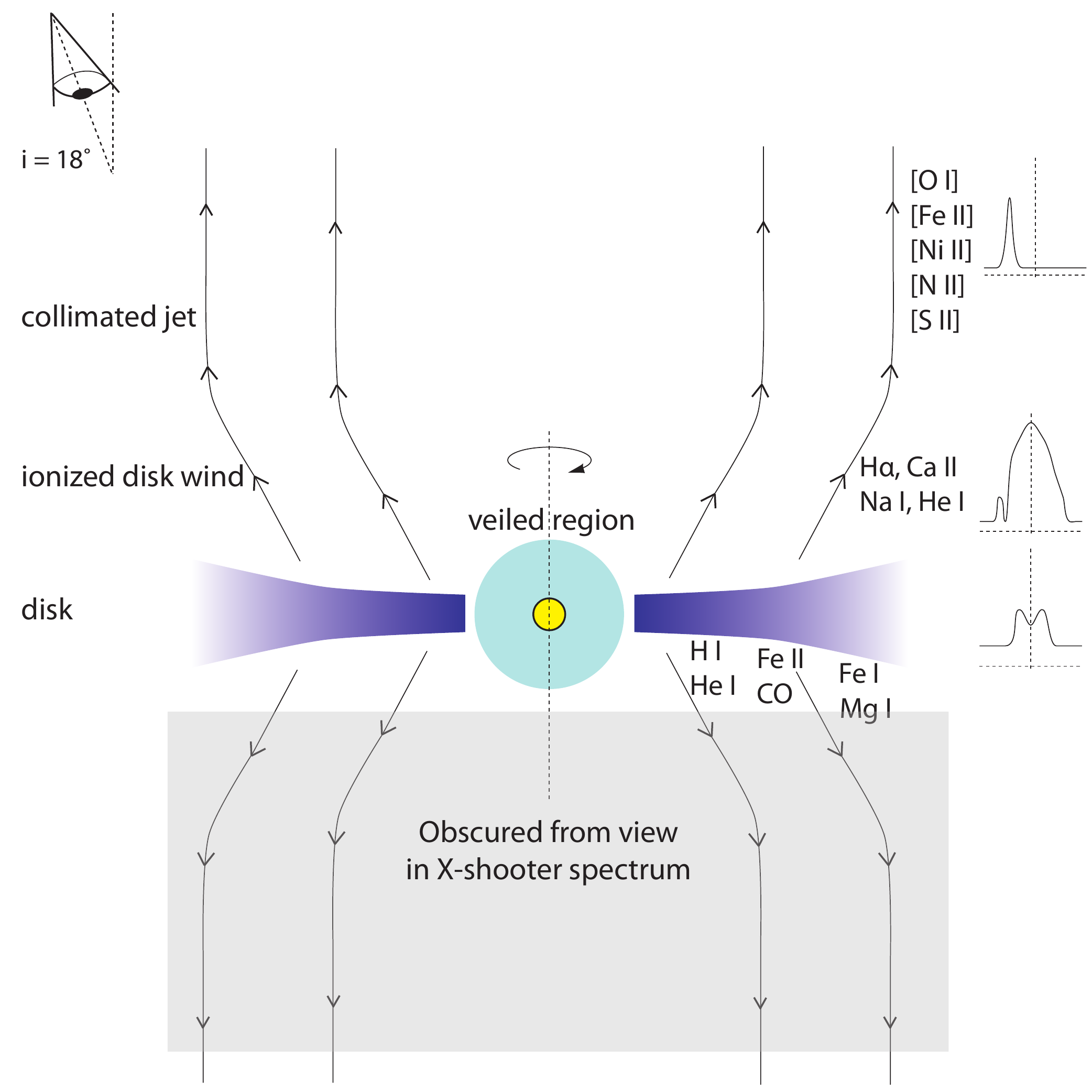}}
\caption{Sketch of the 08576nr292 disk-jet system. The typical line profiles associated with the different components are indicated.}
\label{fig:artistimpression}
\end{figure}

The different line profiles described above and the extended emission morphology show that 08576nr292 is a source with multiple components. A sketch of our interpretation of the system is depicted in Figure~\ref{fig:artistimpression}. The typical line profiles that we associate with the different components are displayed.

The circumstellar disk is the probable origin of the many double-peaked emission lines. The double-peaked morphology likely depends on line strength (and thus the location and extent of the line-forming region), as indicated by the hydrogen spectrum, where only the higher transitions ($3 \rightarrow 10$ and up) show double-peaked emission lines, although not very pronounced. The lower ionization species (Mg~{\sc i}, Fe~{\sc i}) typically have small peak separations, consistent with being formed in the outer part of a Keplerian disk, shielded from the ionizing flux. He~{\sc i} and Fe~{\sc ii} have a larger peak separation, thus presumably formed in a more rapidly rotating inner disk region. The strength of some of the single-peaked emission lines as opposed to the double-peaked lines could point out that they are formed over a more extended region of the Keplerian disk. Additional contributions to these lines may come from an accretion column and an ionized disk wind.

The broad blue-shifted absorption in He~{\sc i} $\lambda$1083~nm is an indicator of outflow (\citealt{Kwan:2010p1267}). We note that absorption over the velocity range up to $-450$~km~s$^{-1}$ is not seen in any other line. The presence of a collimated bipolar jet is unambiguously established by the SINFONI observations. From this we infer that the forbidden emission lines with a radial velocity $\sim 125$~km~s$^{-1}$ are formed in the jet. 

The P-Cygni like profiles of H$\alpha$, Ca~{\sc ii} and Na~{\sc i} are similar to those seen in some accreting Herbig Ae/Be (HAeBe) and ``classical'' T Tauri stars (CTTS, \citealt{Hamann:1992p674, Hamann:1992p661}). Just as in 08576nr292, in some of these stars the Ca~{\sc ii} triplet lines are very strong and their flux ratios indicate saturation. Furthermore, in CTTS, these lines are known tracers of the accretion-infall zone (e.g. \citealt{Muzerolle:1998p796, Muzerolle:2001p788}). 

The coincidence of the blue-shifted absorption features in these lines and the peaks of the forbidden emission lines suggests that they trace the same region. The absorption may take place in the jet or in the disk wind, which would then be the launch region of the jet. In the CTTS models of \cite{Lima:2010p97} a very similar H$\alpha$ profile is produced in an accretion column in combination with an ionized disk wind. 08576nr292 could be accreting in a similar way. 

\begin{figure}[t]
\center{\includegraphics[height=8cm]{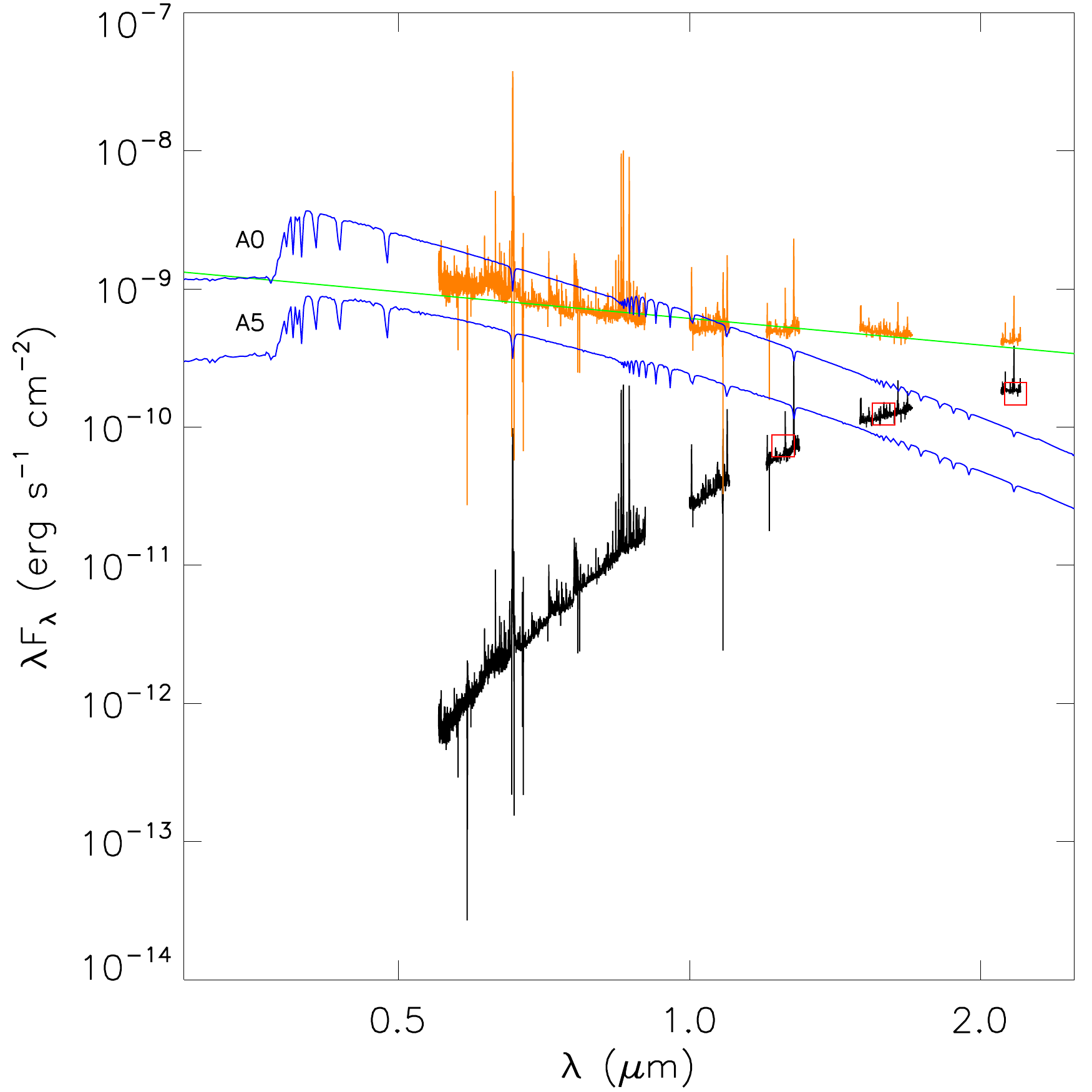}}
\caption{SED as observed by X-shooter (black), de-reddened with $A_V=8$~mag (orange). Furthermore plotted are a power law fitted to the dereddened SED ($\beta=-0.6$, green), \cite{Kurucz1993} models for A0V and A5V stars with ZAMS radii (2.5 and 1.7 $R_{\odot}$) at $d = 0.7$~kpc (blue). Indicated with red squares are the 2MASS fluxes. }
\label{fig:sed}
\end{figure}

\subsection{Evolutionary status and mass}

The presence of a jet and the accretion morphology of the line profiles indicate that 08576nr292 is a YSO with an active accretion disk, contrary to the earlier interpretation that the circumstellar disk is a remnant passive disk (B06). Its mass and accretion rate, however, remain uncertain. The absence of photospheric lines prevents a spectral classification. The shape of the spectral energy distribution (SED) includes information about the mass and luminosity of the source. 

The SED has to be corrected for interstellar extinction (cf. \citealt{Cardelli:1989p1322}), which is a main source of uncertainty. B06 determine $A_V = 12$ mag based on the spectral classification of three O stars in the same cluster, but the reddening is known to vary locally. This is underlined by the fact that the diffuse interstellar bands (DIBs) observed in these O stars (Ellerbroek et al., in prep.) are lacking in 08576nr292. However, the limited visibility of the receding jet lobe suggests that it is not a foreground object. A relatively low extinction could be explained by the fact that the jet has carved out a `hole' in the ambient molecular cloud.

With our current knowledge, we make a refined estimate of $A_V$ assuming that $R_V=3.1$ and $d=0.7$~kpc. The absence of absorption features indicates that the photosphere is heavily veiled due to the accretion. Adopting a power law for the SED ($\lambda F_\lambda \propto \lambda^{\beta}$) we expect $-4/3 < \beta < 0$ in the range 500~--~1000~nm (cf. \citealt{Hillenbrand:1992p1747}). Dereddening the spectrum to match this flat SED leads to $A_V = 8 \pm 1$~mag. The assumption that disk and/or accretion luminosity dominates the SED allows to determine a limit on the spectral type. As can be seen from Fig.~\ref{fig:sed}, this would exclude (PMS equivalents of) stars with spectral type earlier than A0V.

Several remarks should be made. Firstly, the optical SED might not be flat or slightly declining, but increasing due to a cold gas disk. However, the many double-peaked emission lines, and the strong H$\alpha$, suggest that a hot and (partly) ionized disk is present. Secondly, a possible underestimate of the distance would imply a larger upper bound on the stellar luminosity. 

\cite{Hartigan:1995p1041} derive empirical relations between the line fluxes of optical forbidden lines and the mass-loss rate through the jet for CTTS. Applying these relations to the de-reddened spectrum of 08576nr292, we determine a jet mass-loss rate in the order of 10$^{-8}$ -- 10$^{-7}$~$M_\odot$~yr$^{-1}$. The same study suggests that the mass accretion rate is typically 10$^2$ times the mass-loss rate, implying an accretion rate of 10$^{-5}$ -- 10$^{-6}$~$M_\odot$~yr$^{-1}$. Only a few intermediate-mass stars have been observed with such high rates (\citealt{GarciaLopez:2006}). In this case one would expect the disk to be optically thick up to distances close to the stellar surface (up to a few stellar radii, cf. \citealt{Hillenbrand:1992p1747}). This would be consistent with the disk dominating the SED up to optical wavelengths.

These considerations suggest that 08576nr292 is a HAeBe-like, Lada Class II source, undergoing active accretion. Because of its brightness, relatively low extinction and low inclination angle, 08576nr292 provides a unique opportunity to study an active disk-jet system in great detail. Further X-shooter observations and modelling of the jet are planned, aimed to further explore the dynamics of this system. 

\begin{acknowledgements}
LE is funded by NOVA. EP is partially funded by Consolider-Ingenio 2010 Program CSD2006-00070. The authors thank Mario van den Ancker for useful discussions.
\end{acknowledgements}

\newpage

\begin{center}
\vspace*{1cm}

{\Large {\bf Appendix: 

X-shooter spectrum of 08576nr292}}
\vspace{1cm}

Observed 23-02-2010 UT 4h50

Not corrected for telluric absorption lines
\end{center}

\clearpage

\begin{figure*}[t]
  \center{\includegraphics[width=18cm]{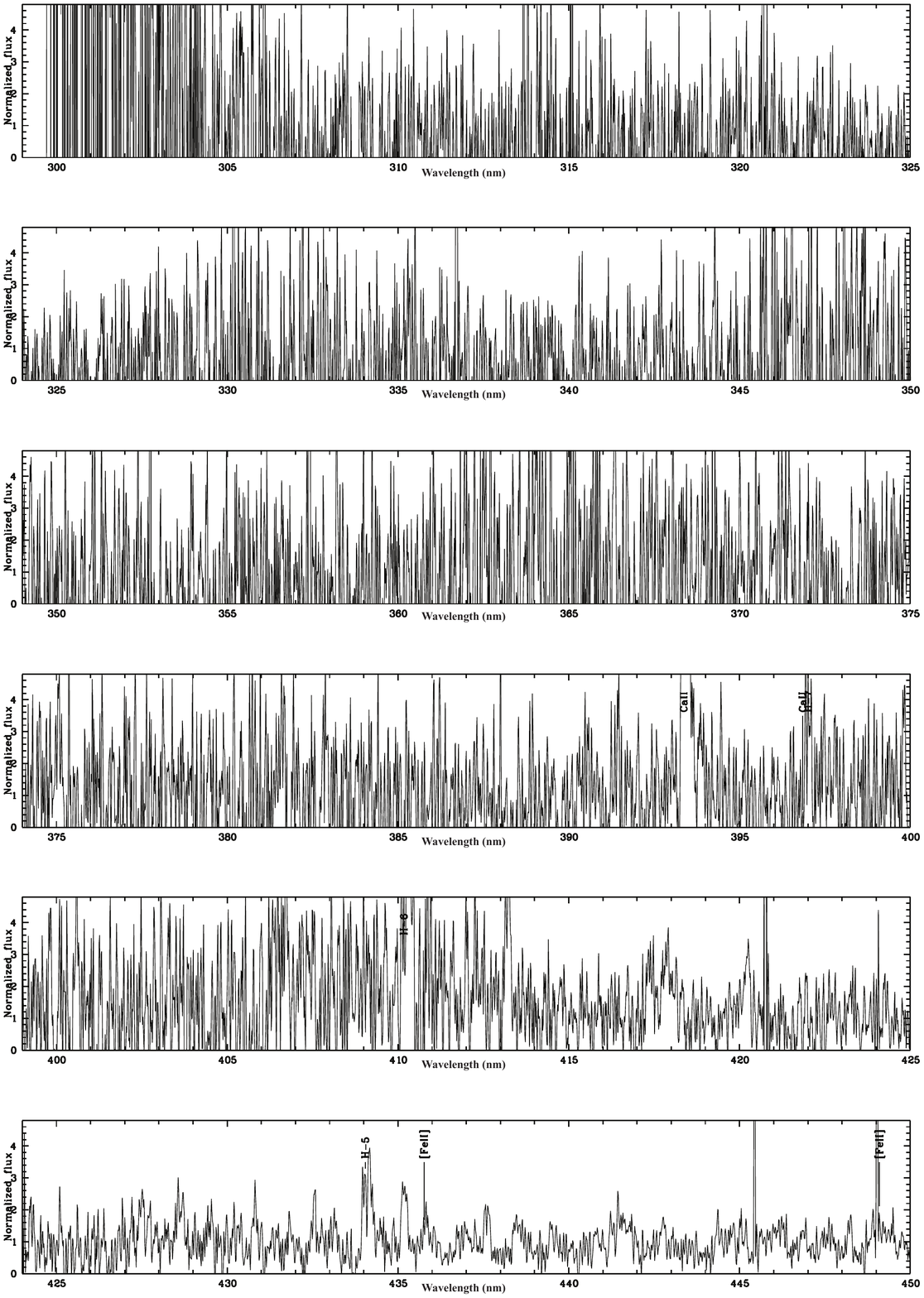}}
\end{figure*}
\clearpage

\begin{figure*}[t]
  \center{\includegraphics[width=18cm]{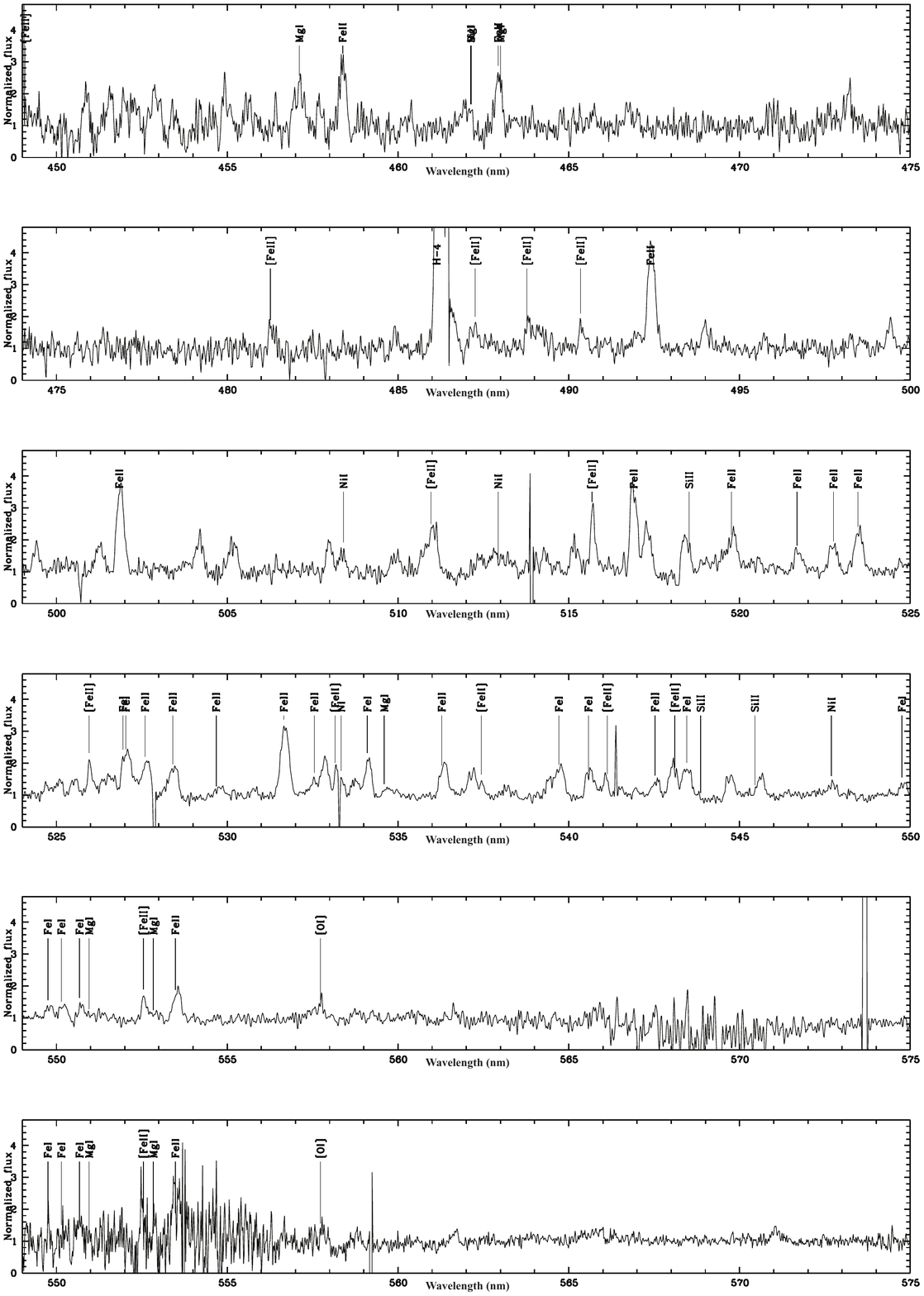}}
\end{figure*}
\clearpage

\begin{figure*}[t]
  \center{\includegraphics[width=18cm]{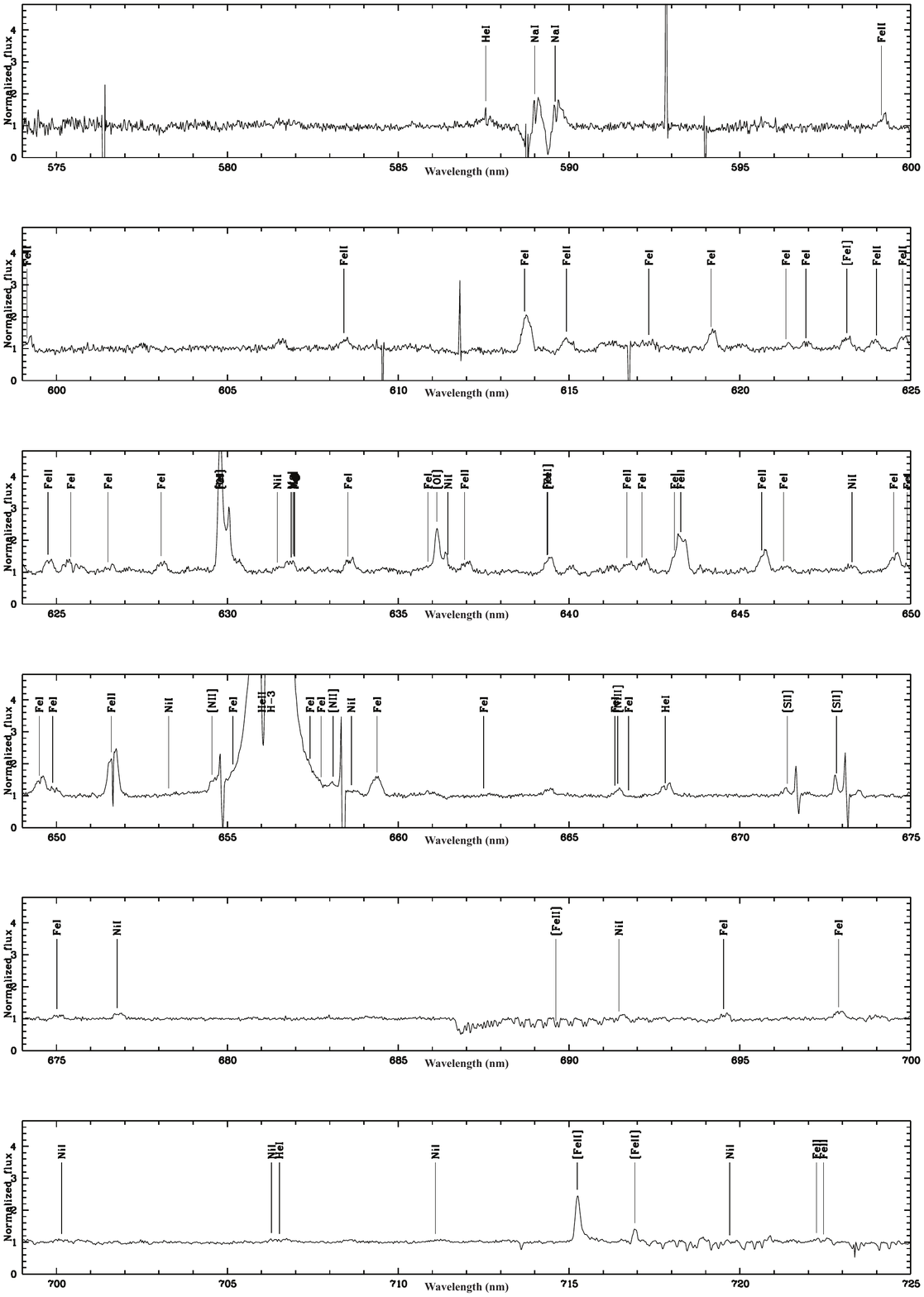}}
\end{figure*}
\clearpage

\begin{figure*}[t]
  \center{\includegraphics[width=18cm]{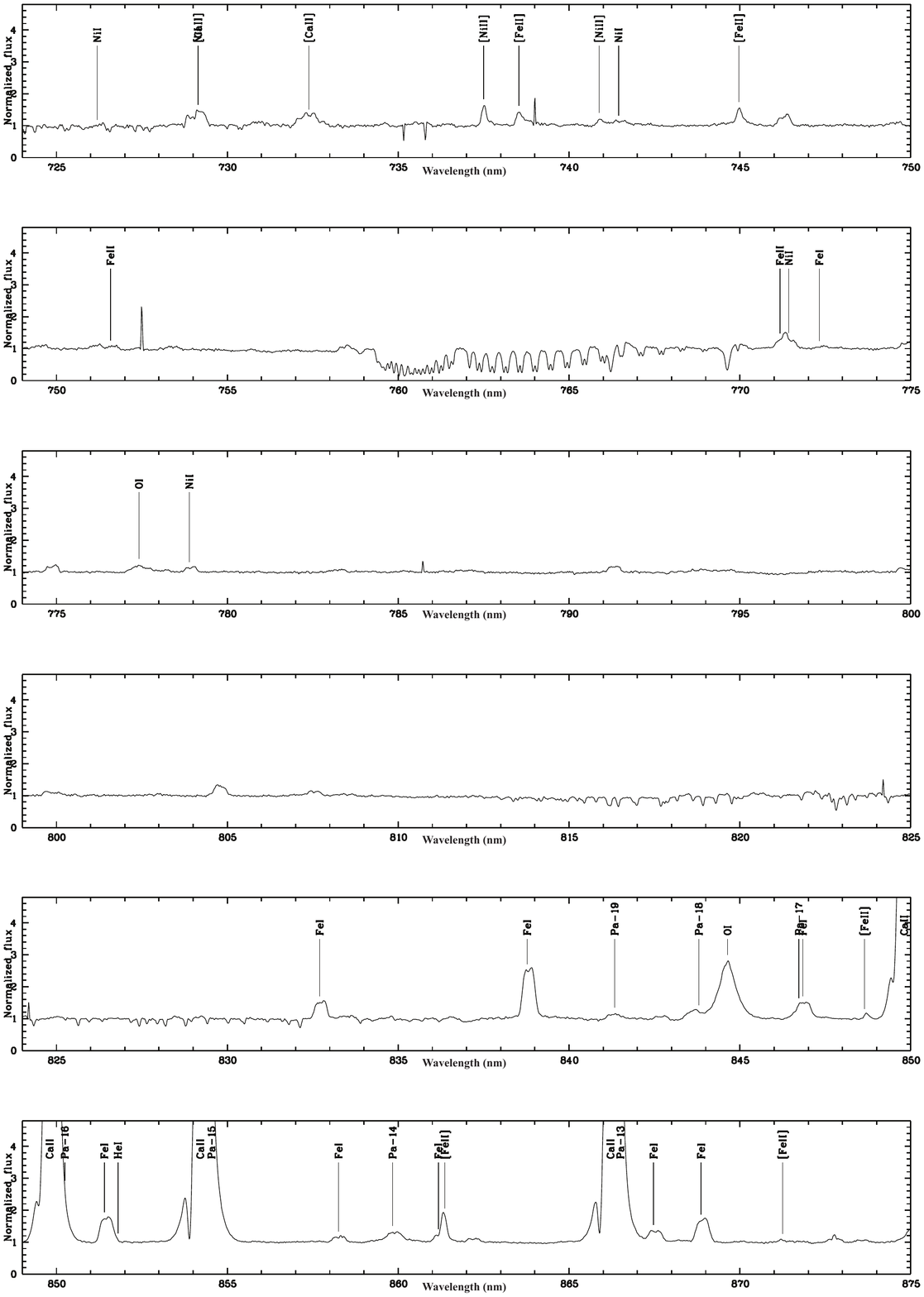}}
\end{figure*}
\clearpage

\begin{figure*}[t]
  \center{\includegraphics[width=18cm]{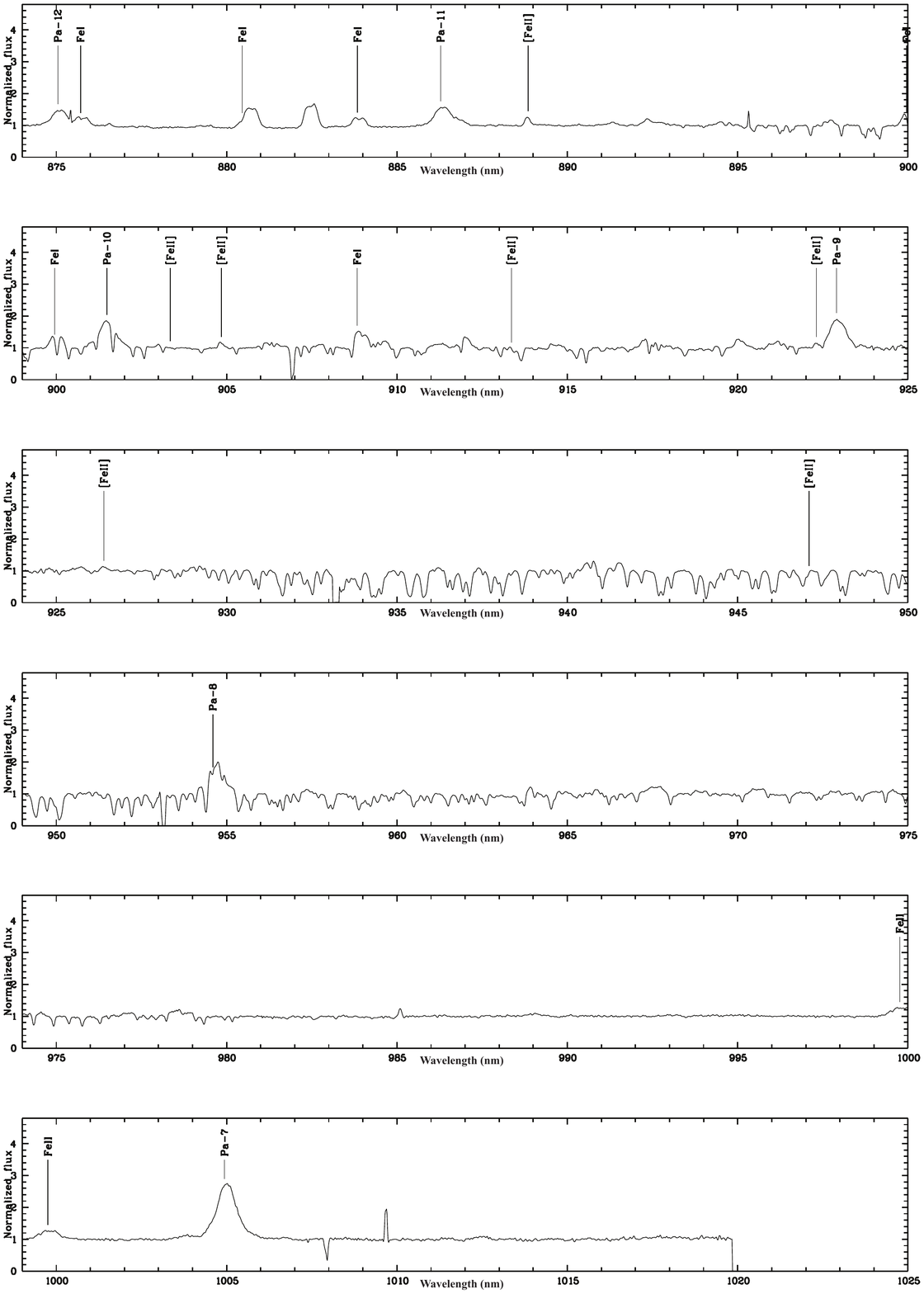}}
\end{figure*}
\clearpage

\begin{figure*}[t]
  \center{\includegraphics[width=18cm]{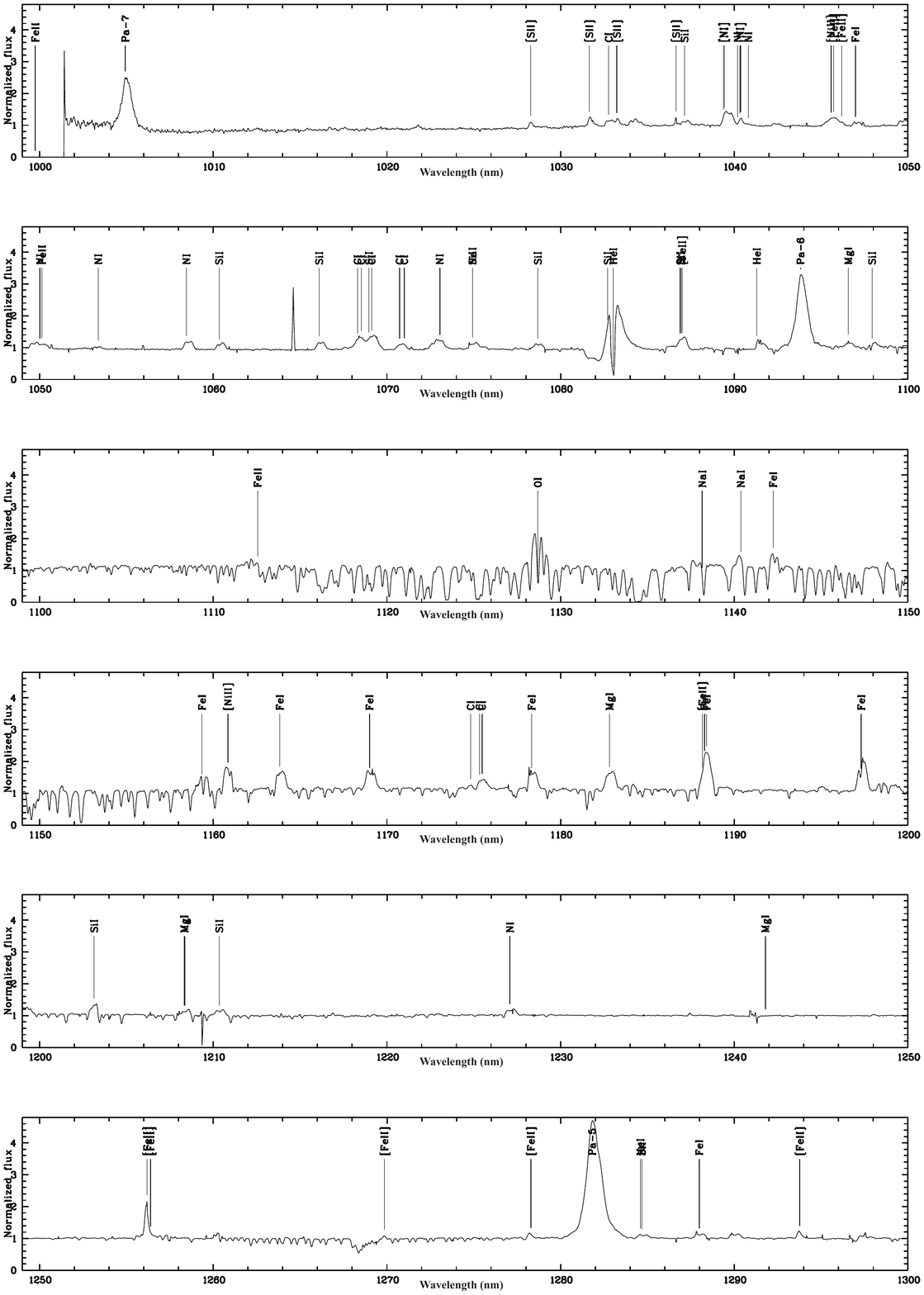}}
\end{figure*}
\clearpage

\begin{figure*}[t]
  \center{\includegraphics[width=18cm]{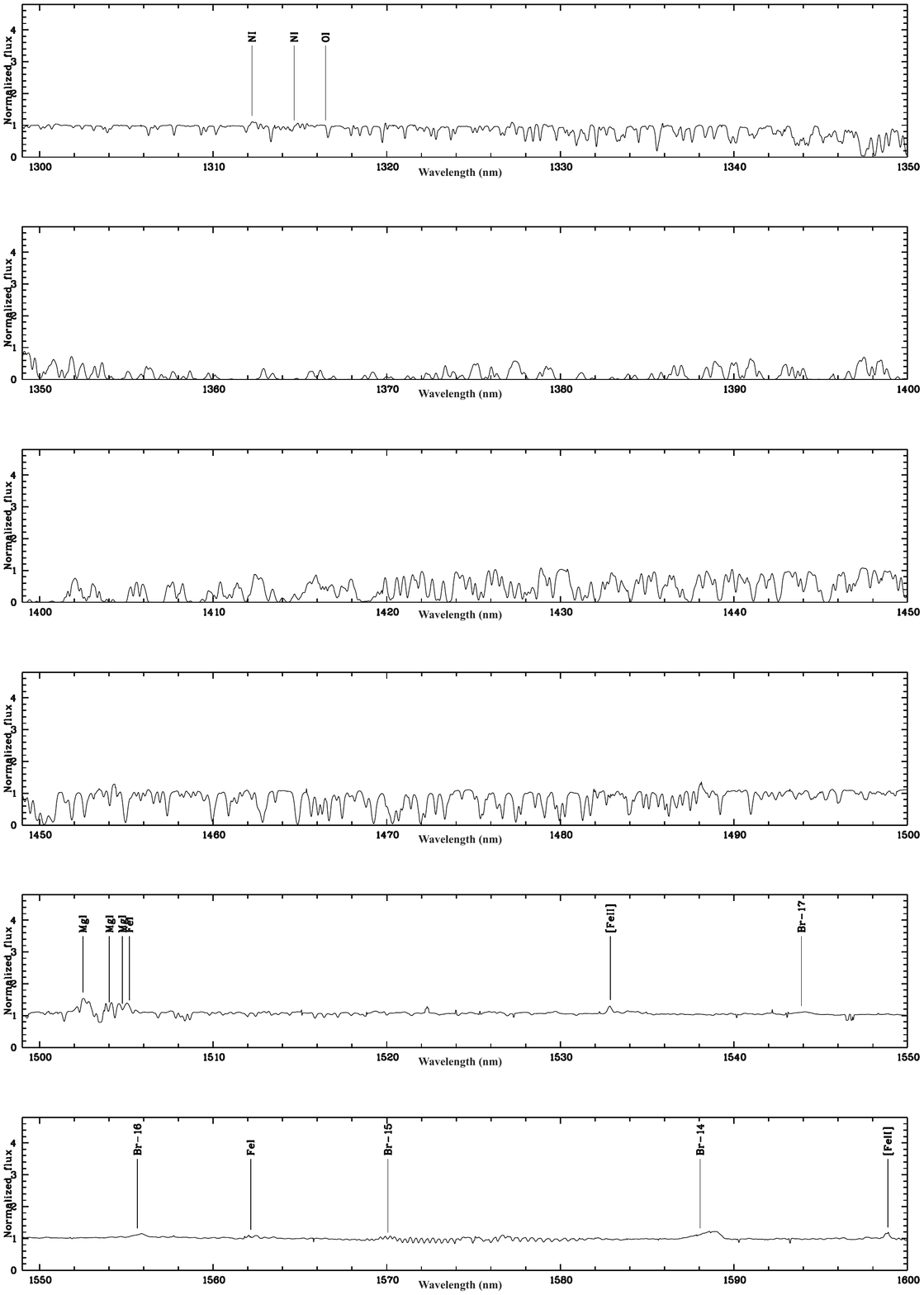}}
\end{figure*}
\clearpage

\begin{figure*}[t]
  \center{\includegraphics[width=18cm]{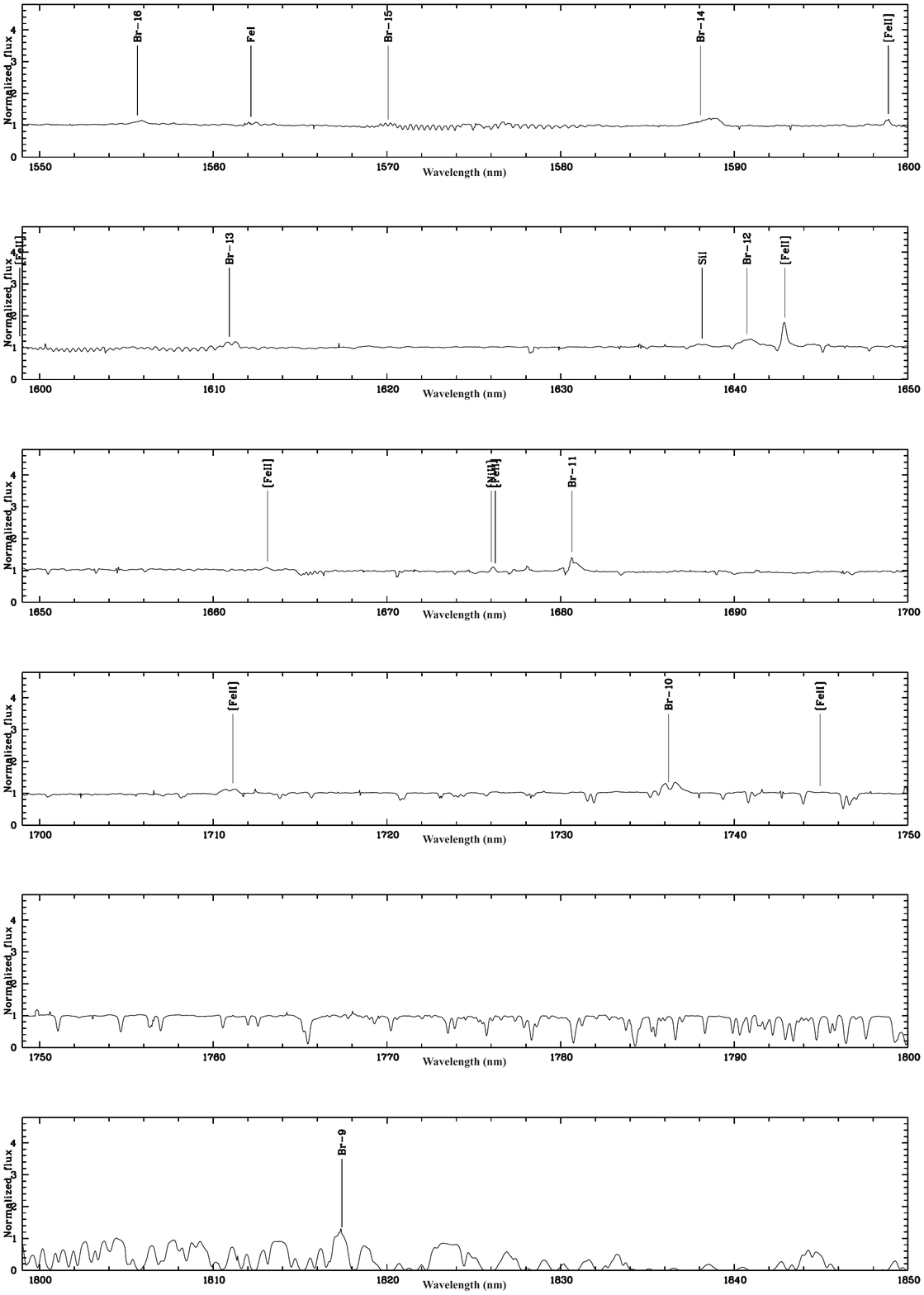}}
\end{figure*}
\clearpage

\begin{figure*}[t]
  \center{\includegraphics[width=18cm]{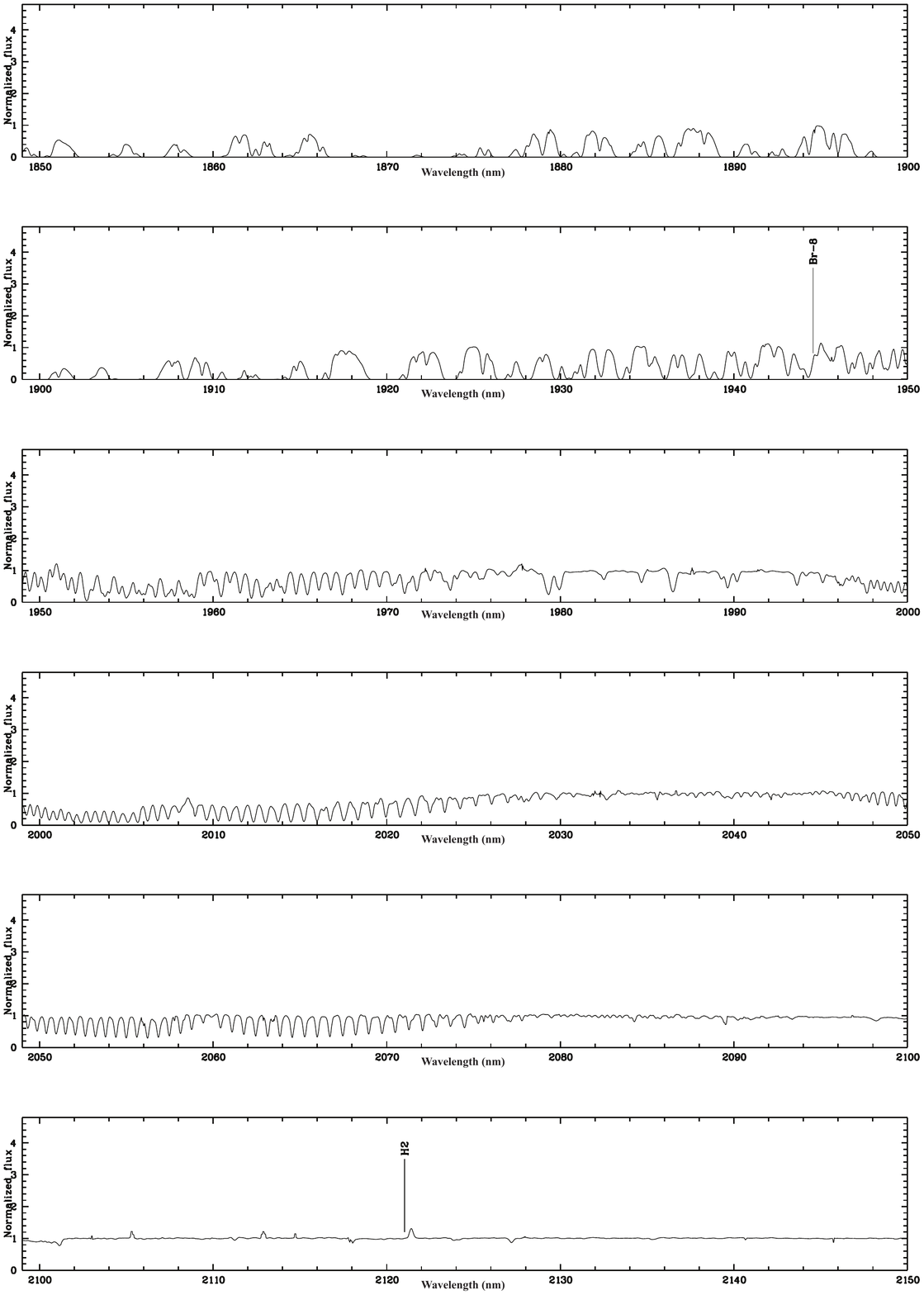}}
\end{figure*}
\clearpage

\begin{figure*}[t]
  \center{\includegraphics[width=18cm]{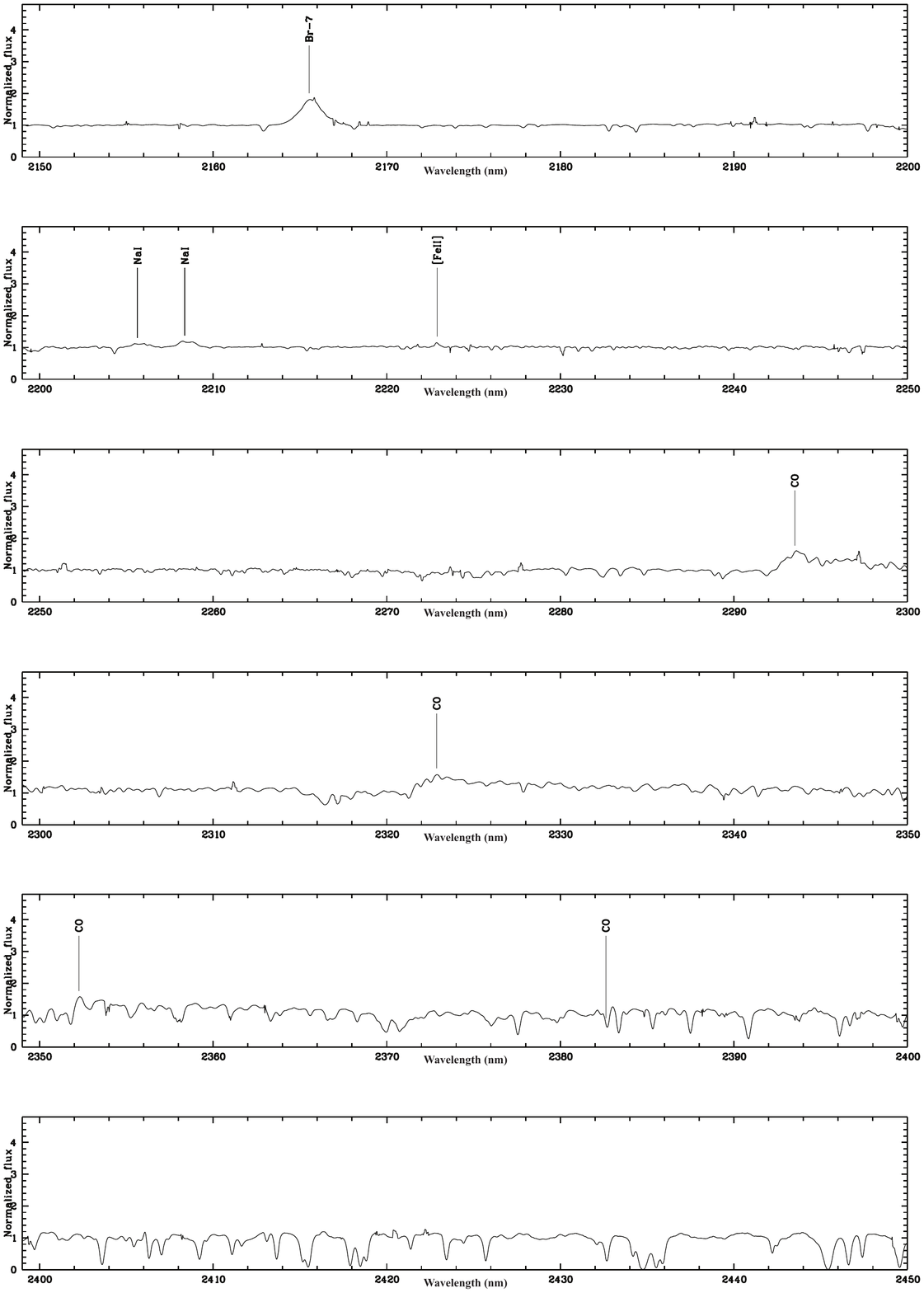}}
\end{figure*}
\clearpage

 \end{document}